%% file: NN4toricGeoFINAL.tex
\documentclass[aps,amsmath,amssymb,preprintnumbers,groupedaddress,twocolumn]{revtex4}
\usepackage{graphicx,psfrag,hyperref}
\usepackage{amsmath,bbm,array,amsfonts,graphicx,wrapfig,lscape,float,mathtools,multirow,longtable}
\usepackage[dvipsnames]{xcolor}

\input{pref}

\newcommand{\req}[1]{(\ref{#1})}
\newcommand{\ie}{{\it i.e.}}

\begin{document}

\preprint{CERN-TH-2017-128}
\preprint{UUITP-17/17}
\title{Machine Learning of Calabi-Yau Volumes}

\author{Daniel Krefl$^{a}$ and Rak-Kyeong Seong$^{b}$}

\affiliation{\it ${}^{a}$ 
Theoretical Physics Department, CERN, Geneva 23, CH-1211 Switzerland\\ 
\it ${}^{b}$ 
Department of Physics and Astronomy, Uppsala University, SE-751 08 Uppsala, Sweden
}

\begin{abstract}
We employ machine learning techniques to investigate the volume minimum of Sasaki-Einstein base manifolds of non-compact toric Calabi-Yau 3-folds. 
We find that the minimum volume can be approximated via a second order multiple linear regression on standard topological quantities obtained from the corresponding toric diagram. 
The approximation improves further after invoking a convolutional neural network with the full toric diagram of the Calabi-Yau 3-folds as the input. 
We are thereby able to circumvent any minimization procedure that was previously necessary and find an explicit mapping between the minimum volume and the topological quantities of the toric diagram. 
Under the AdS/CFT correspondence, the minimum volumes of Sasaki-Einstein manifolds correspond to central charges of a class of $4d$ $\mathcal{N}=1$ superconformal field theories.
We therefore find empirical evidence for a function that gives values of central charges without the usual extremization procedure. 
\end{abstract} 
\maketitle
\noindent

\section{Introduction}

In recent years machine learning has become a cornerstone for many fields of science and it has been adopted more and more as a valuable toolbox. 
Machine learning has attracted much interest due to significant theoretical progress and due to increased availability of large amounts of data, computing power (GPUs) and easy to use software implementations of standard machine learning techniques.

Despite these developments, applications of machine learning techniques to mathematical physics have been limited  to our knowledge. 
One of the reasons for this is that machine learning aims to empirically approximate the underlying probability density function of a given dataset.
Making use of machine learning to identify hidden structures in datasets, which teach us about new phenomena in string theory and mathematics, has not been systematically considered before.

This work aims to change the status quo and to provide evidence that machine learning can be used to discover hidden structures in large classes of gauge theories that are studied in theoretical physics as well as large classes of geometries that are studied in mathematics. 
Importantly, we illustrate that machine learning does not just provide an approximation of known functional relationships between
physically and mathematically significant quantities, but also leads to discoveries of new functional relationships.

In particular, we concentrate on a class of $4d$ $\mathcal{N}=1$ supersymmetric gauge theories that live on the worldvolume of a stack of D3-branes probing toric Calabi-Yau 3-folds, characterized by convex lattice polygons known as toric diagrams \cite{Klebanov:1998hh,Hanany:1997tb,Hanany:1998it,Franco:2005rj}. These theories are expected to flow at low energies to a superconformal fixed point. 
\\

From a machine learning perspective, this work studies the minimum volumes of Sasaki-Einstein 5-manifolds. These are the base manifolds of the probed toric Calabi-Yau 3-folds.
The minimum volume is of particular interest because under the AdS/CFT correspondence, it is expected to be related to the maximized $a$-function that gives the central charge of the $4d$ $\mathcal{N}=1$ superconformal field theory \cite{Gubser:1998vd,Henningson:1998gx,Intriligator:2003jj,Butti:2005vn,Butti:2005ps}.

Using a large dataset of toric Calabi-Yau 3-folds, our aim is to train a machine learning model in such a way that it approximates a functional relationship between topological quantities of the toric Calabi-Yau 3-fold and the minimum volume of the Sasaki-Einstein base manifold.
Such a functional relation would be of great use because it would circumvent the standard volume minimization procedure and highlight a direct relationship between topological quantities of the toric Calabi-Yau geometries and the central charges of the $4d$ superconformal field theories.

We ask whether for a given dataset consisting of minimal volumes $V_{min}$ and topological quantities $\mathcal T$ of the corresponding Calabi-Yau 3-folds, we can approximate from the data a mapping $F$ such that
$$
V_{min} \sim F(\mathcal T)\,. 
$$
Here, we use machine learning to model $F$.
Mainly, we will consider three different machine learning models. These are a modification of usual linear regression, a convolutional neural network (CNN) and also a combination of both.
In detail, a CNN model is a feed-forward neural network which includes additional convolutional layers. 

We refer the reader to \cite{LeCun2015} for a basic introduction on CNN models and \cite{Schmidthuber} for a comprehensive reference list.
\\

\section{Background}\label{sback}

We concentrate on non-compact Calabi-Yau 3-folds $\mathcal{X}$ that are realized as affine cones over a complex base $X$.
In particular, we focus on a special subclass of $\mathcal{X}$ where the base is a toric variety $X(\Delta)$, which is defined by a convex lattice polygon $\Delta$ known as the toric diagram.
$\mathcal{X}$ can also be thought of as the real cone over a compact, smooth Sasaki-Einstein 5-manifold $Y$, whose metrics have been studied extensively for various classes of toric Calabi-Yau 3-folds.
The K\"ahler metric of $\mathcal{X}$ has the form
\beal{es1}
\ud s^2 (\mathcal{X}) = \ud r^2 + r^2 \ud s^2 (Y) ~,~
\eea
where $Y=\mathcal{X}|_{r=1}$.
\\

When we talk of Calabi-Yau volumes, we actually refer to the volume function of the Sasaki-Einstein base $Y$, which takes the form
\beal{es2}
\text{vol}[Y] = \int_Y \ud \mu =  \int_{r\leq 1} \omega^3
~,~
\eea
where $\ud \mu$ is the Riemannian measure on the cone $\mathcal{X}$.
$\omega$ is the K\"ahler form and is given by
\beal{es3}
\omega = - \frac{1}{2} \ud(r^2 \eta) = \frac{1}{2} i \partial \overline{\partial} r^2 ~,~
\eea
where $\eta$ is a global one-form on $Y$. 
Normalized under the volume of $S^5$, we denote the volume function of $Y$ as
\beal{es4}
V(b;Y) := \frac{\text{vol}[Y]}{\text{vol}[S^5]} ~.~
\eea
Note that $V(b;Y)$ is an algebraic number and is expressed in terms of Reeb vector components $b_{i=1,\dots,3}$, where $b_3=3$ \cite{Martelli:2005tp,Martelli:2006yb}.

For a given projective variety $X$, realized as an affine variety in $\mathbb{C}^k$, the Hilbert series is the generating function for the dimension of the graded pieces of the coordinate ring $\mathbb{C}[x_1,\dots,x_k]/\langle f_i\rangle$, where $f_i$ are the defining polynomials of $X$. 
The Hilbert series of $X$ takes the form of a rational function with the expansion,
\beal{es5}
g(t;\mathcal{X}) = \sum_{i=0}^{\infty} \text{dim}_{\mathbb{C}}(X_i) t^i ~,~
\eea
where the $i$th graded piece $X_i$ can be thought of as the number of algebraically independent degree $i$ polynomials on the variety $X$, with $t$ keeping track of the degree $i$.

\begin{figure}
  \includegraphics[scale=0.38]{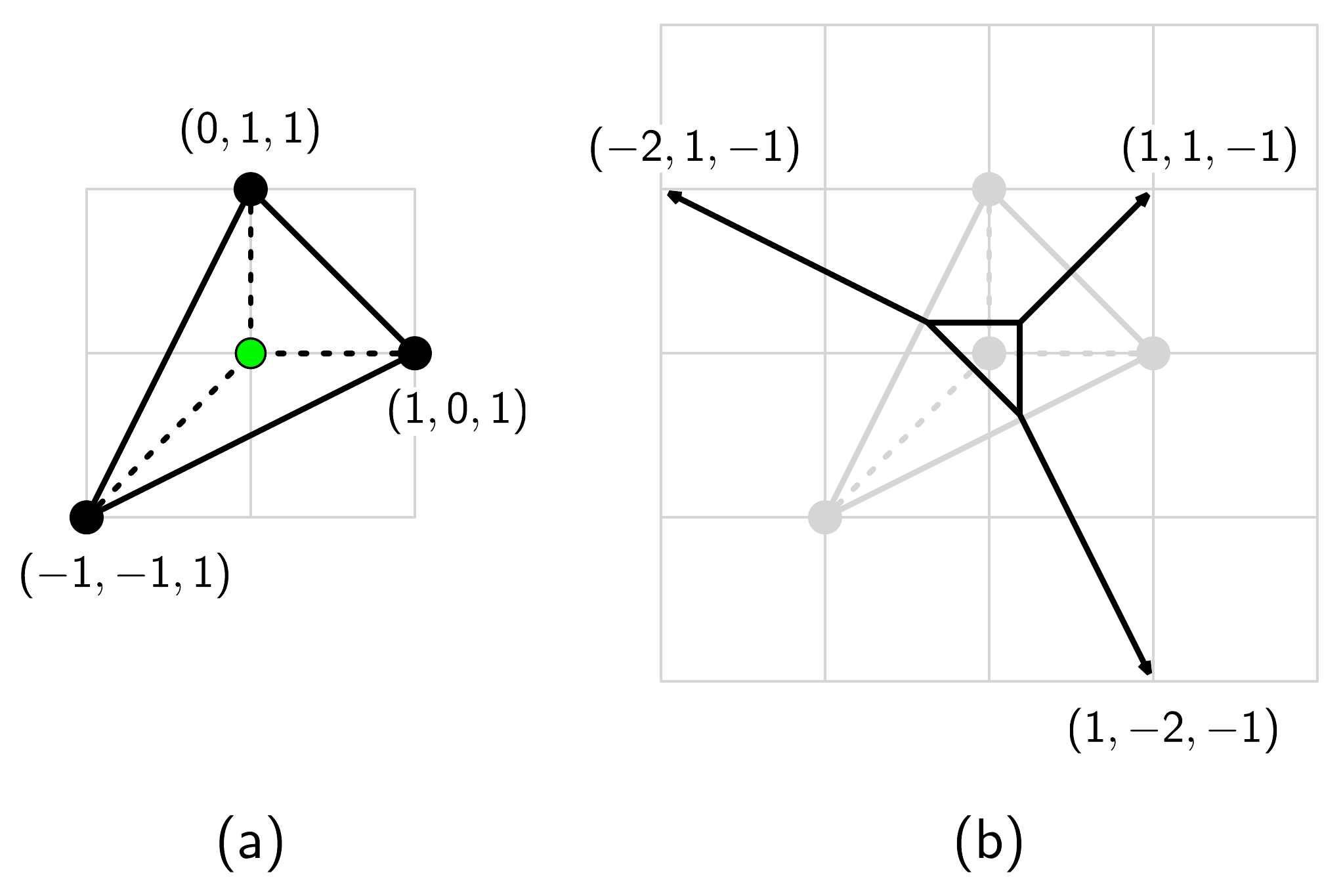}
  \caption{
  (a) shows the toric diagram for $\mathbb{C}^3/\mathbb{Z}_3$ and the corresponding ideal triangulation into unit area triangles. (b) is the corresponding dual web-diagram with normal vectors to each boundary edges of the triangulation. Notice that we have 3-vectors, given that the original toric diagram is on a plane at height 1. 
}
  \label{ftorictriang}
\end{figure}

When $X$ is a toric variety defined by a convex lattice polygon $\Delta$, the Hilbert series for $X(\Delta)$ and the corresponding Calabi-Yau cone $\mathcal{X}$ can be obtained from the ideal triangulation of the toric diagram $\Delta$,
\beal{es10}
g(t_i; \mathcal{X}) = \sum_{i=1}^{r} \prod_{j=1}^{n} (1-\vec{t}^{~\vec{u}_{i,j}})^{-1} ~,~
\eea
where the index $i=1,\dots,r$ runs over the unit triangles in the ideal triangulation and $j=1,2,3$ runs over the boundary edges of each such triangle \cite{Martelli:2005tp,Benvenuti:2006qr}. 
$\vec{u}_{i,j}$ is a 3-dimensional outer normal to the edge $j$ of the associated unit triangle $i$, where $\vec{t}^{~\vec{u}_{i,j}}= \prod_{a}^3 t_a^{u_{i,j}(a)}$. 
Note that we are dealing with 3-vectors because the 2-dimensional toric diagram is on a plane at height 1.
Using \eref{es10}, the Hilbert series for $\mathbb{C}^3/\mathbb{Z}_3$, whose toric diagram is shown in \fref{ftorictriang}, can be obtained as follows
\beal{es11}
&&
g(t_i; \mathbb{C}^3/\mathbb{Z}_3) 
=
\frac{1}{(1-t_2)(1-t_1^{-1}t_2)(1-t_1 t_2^{-2} t_3^{-1})} 
\nn\\
&&
\hspace{1.5cm}
+ \frac{1}{(1-t_1)(1-t_1 t_2^{-1})(1-t_1^{-2} t_2 t_3)} 
\nn\\
&&
\hspace{1.5cm}
+ \frac{1}{(1-t_1^{-1})(1-t_2^{-1})(1-t_1 t_2 t_3^{-1})} 
~.~
\eea

The volume function can be derived directly from the Hilbert series of $\mathcal{X}$ following the limit 
\beal{es15}
V(b_i; Y) = \lim_{\mu\rightarrow 0} \mu^3 g(t_i = \exp[-\mu b_i]; \mathcal{X}) ~.~
\eea
The leading order in $\mu$ picked up by the above limit from the expansion of the Hilbert series was shown in \cite{Martelli:2005tp,Martelli:2006yb} to be directly related to the volume of the Sasaki-Einstein base $Y=\mathcal{X}|_{r=1}$. 
For the $\mathbb{C}^3/\mathbb{Z}_3$ example, the volume function takes the form
\beal{es16}
&&
V(b_i; \mathbb{C}^3/\mathbb{Z}_3) = 
\nn\\
&&
\frac{-9}{(b_1-2b_2 -3)(-2 b_1+b_2 -3)(b_1 +b_2 - 3)} 
,~
\eea
where $b_3=3$.

\section{Volume Minimization and the AdS/CFT correspondence}

The worldvolume theory on a stack of D3-branes probing Calabi-Yau 3-folds $\mathcal{X}$ is a $4d$ $\mathcal{N}=1$ supersymmetric gauge theory. 
It is expected that these theories flow at low energies to a superconformal fixed point.
The superconformal R-charges of the theory are determined by a procedure known as $a$-maximization \cite{Intriligator:2003jj,Butti:2005vn,Butti:2005ps}, which involves the maximization of the $a$-charge,
\beal{es20}
a(R; Y) = \frac{3}{32} (3\text{Tr} R^3 - \text{Tr} R) ~.~
\eea
$a$-maximization gives the value of the central charge at the conformal fixed point, which is by the AdS/CFT correspondence related to the volume minimum of the corresponding Sasaki-Einstein 5-manifold \cite{Gubser:1998vd,Henningson:1998gx} under
\beal{es21}
a(R; Y) = \frac{\pi^3 N^2}{4 V(R;Y)} ~,~
\eea
where the R-charge $R$ can be expressed in terms of Reeb vector components $b_i$. 
In other words, computing the minimum volume,
\beal{es22}
V_{min} = \text{min}_{b_i|b_3=3} V(b_i; Y) ~,~
\eea
is equivalent under \eref{es21} to computing the maximized value of $a(R;Y)$, which is the central charge of the $4d$ $\mathcal{N}=1$ superconformal field theory.

\section{Data}\label{sdata}

Our aim is to train a neural network to compute the volume minimum directly as a function of toric data, circumventing the minimization procedure that has been so far necessary. 
The available input data for the machine learning models for a given toric Calabi-Yau 3-fold takes the following form
\beal{classicalFeatures0}
(y,\mathcal{T})\,,\,\,\,\,
\mathcal{T}=(f_1,f_2,f_3,\mathcal D)
\eea
where $y=1/V_{min}$ is the target inverse minimum volume, and $f_i$ are the three features
\beal{classicalFeatures}
f_1= I\,,\,\,\,\, f_2=E \,,\,\,\,\, f_3=V\,,
\eea
with $I$ being the number of internal lattice points, $E$ being the number of perimeter points and $V$ being the number of extremal corner points of the convex lattice polygon representing the toric diagram. 
Note that $2f_2 - 4$ is the Euler number of the corresponding toric variety \cite{He:2017gam}.
In addition, we include the toric diagram itself as a square matrix $\mathcal D$, consisting of $0,1$ entries, where an entry of $1$ indicates the presence of an extremal vertex of the lattice polygon. 

\begin{figure}
  \includegraphics[scale=0.35]{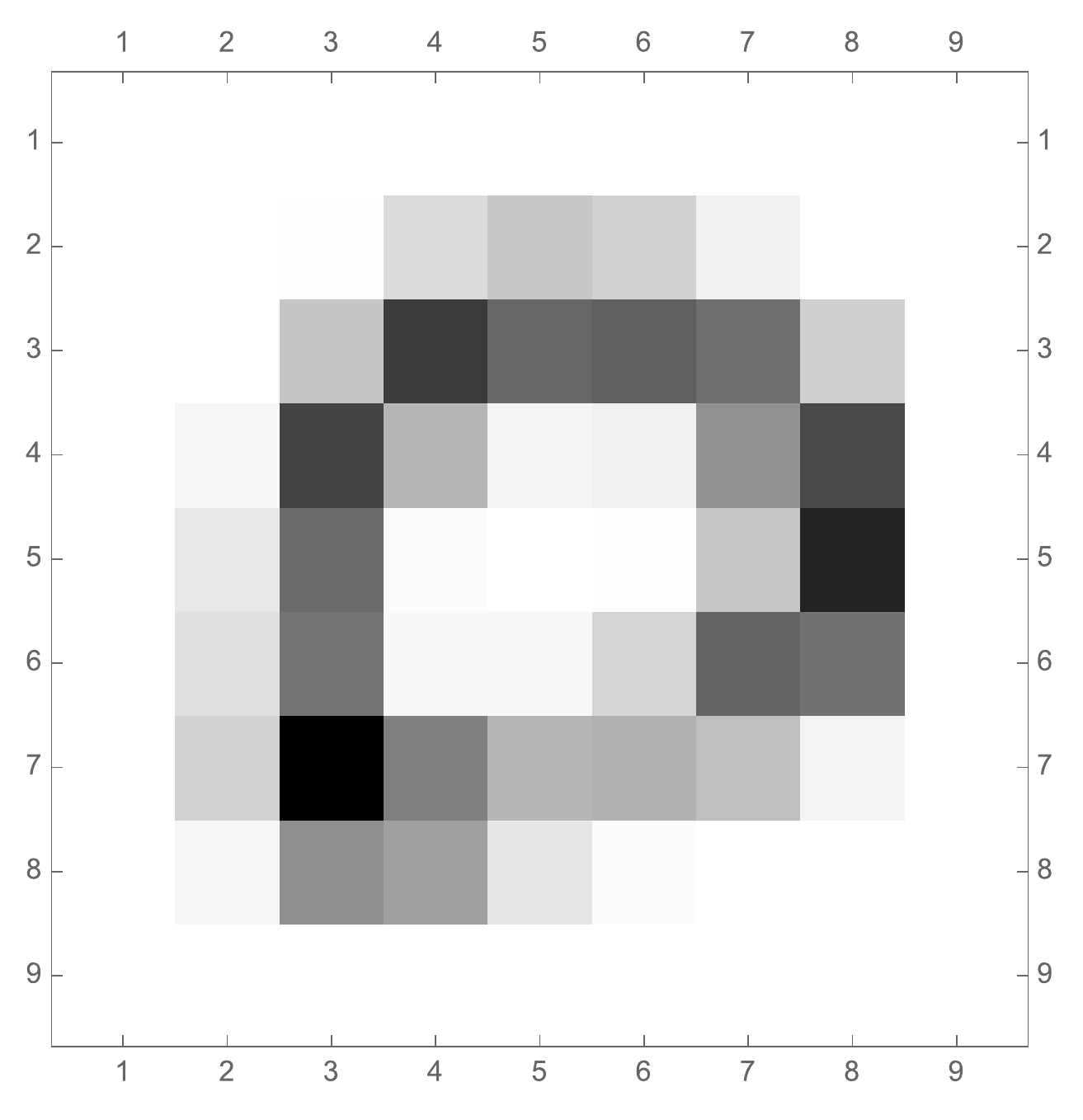}
  \caption{
  The distribution of extremal vertices of toric diagrams for the set of 15,151 distinct toric Calabi-Yau 3-folds that are used as train and test sets for our machine learning models. 
}
  \label{ftoricdiadistr}
\end{figure}

We generate a class of toric Calabi-Yau's whose toric diagrams originate from the toric diagram of the orbifold of the conifold of the form $\mathcal{C}/\mathbb{Z}_5\times \mathbb{Z}_5$, which is a lattice square with side-length 5. 
By consecutively cutting corners of this toric diagram, we generate 187,389 distinct toric diagrams. 
However, this set of toric diagrams exhibits a remaining $GL(2,\mathbb{Z})$ redundancy and hence certain toric diagrams from this set can be related to the same toric Calabi-Yau 3-fold. 
We therefore remove the $GL(2,\mathbb{Z})$ redundancy and further reduce the number of toric diagrams down to 15,151, which now establishes a set of distinct toric Calabi-Yau 3-folds. 
Using the integer rounded centroid of the convex lattice polygons, we re-center the toric diagrams.
All the 15,151 re-centered toric diagrams then fit into a 7x7 lattice square, which we further embed into a 9x9 lattice square. Accordingly, $\mathcal{D}$ for our dataset is a 9x9 integer matrix with entries $0,1$ .
In \fref{ftoricdiadistr}, we illustrate the distribution of the extremal vertices of all the 15,151 toric diagrams we use for our analysis.

\begin{figure}
  \includegraphics[scale=0.25]{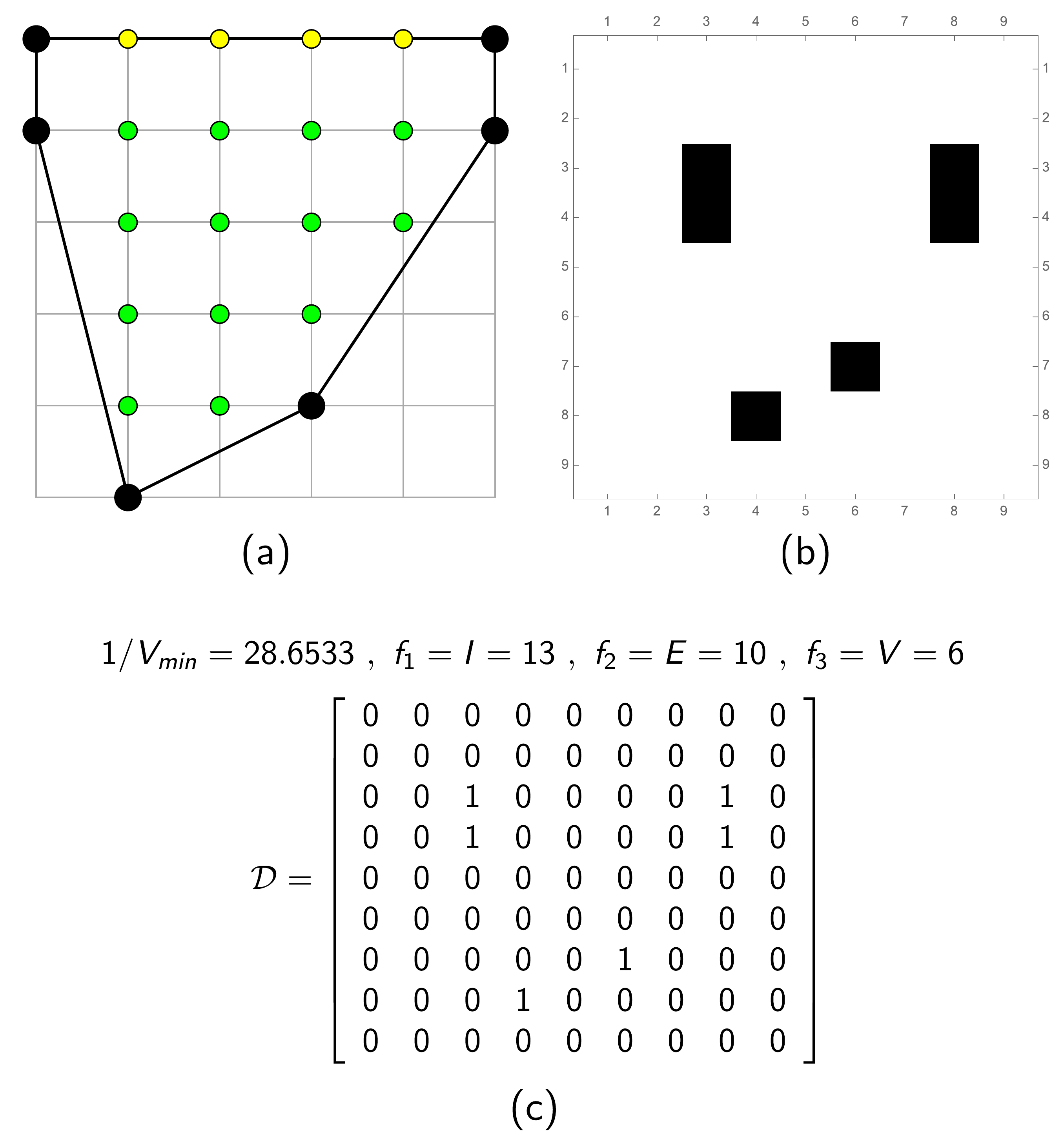}
  \caption{
An example of a data entry vector for a particular toric Calabi-Yau 3-fold in our dataset. (a) shows the toric diagram obtained by cutting corners of a 5x5 lattice square and (b) shows the corresponding extremal lattice points of the toric diagram embedded in a 9x9 input matrix. The full data vector for this toric Calabi-Yau 3-fold is summarized in (c).
}
  \label{ftoricdatasetexample}
\end{figure}

Using Hilbert series we compute the volume function $V(b_i;Y)$ for our dataset and minimize them to obtain $V_{min}$. 
Given the entire dataset with the minimized volumes, we identify 4 cases where the value of  $y=1/V_{min}$ is much larger than the remaining dataset. 
In order to keep a non-distorted dataset, we remove these 4 cases ending up with a dataset of size 15,147.
We also note that the minimum volumes are algebraic numbers, which can be irrational, and that we round the actual values for the machine learning model to 4 decimal points.
This gives us 797 distinct values for the minimum volume under the chosen numerical precision, corresponding to the 15,147 distinct toric Calabi-Yau 3-folds.
Finally, for each toric diagram in our set of Calabi-Yau 3-folds, we identify the features $f_i$ that lead us to 15,147 input data of the form shown in \eref{classicalFeatures0}. An example is illustrated in \fref{ftoricdatasetexample}.

As neural networks easily overfit, meaning that they tend to memorize the specific training set instead of learning the underlying hidden structure, we have to be very careful in data preparation and usage in order not to fool ourselves. 
We here follow the common approach to split the data into an independent train (75\%) and test (25\%) set, where the machine learning models are only trained on the train set. 
We do not use an additional verification set, as we will not perform extensive hyper-parameter optimization. 
It is important to note that we have constructed our dataset of 15,147 toric Calabi-Yau 3-folds in such a way that a split will give $GL(2,\mathbb{Z})$ independent train and test sets.

\section{Multiple Linear Regression}

\begin{figure}
  \includegraphics[scale=0.55]{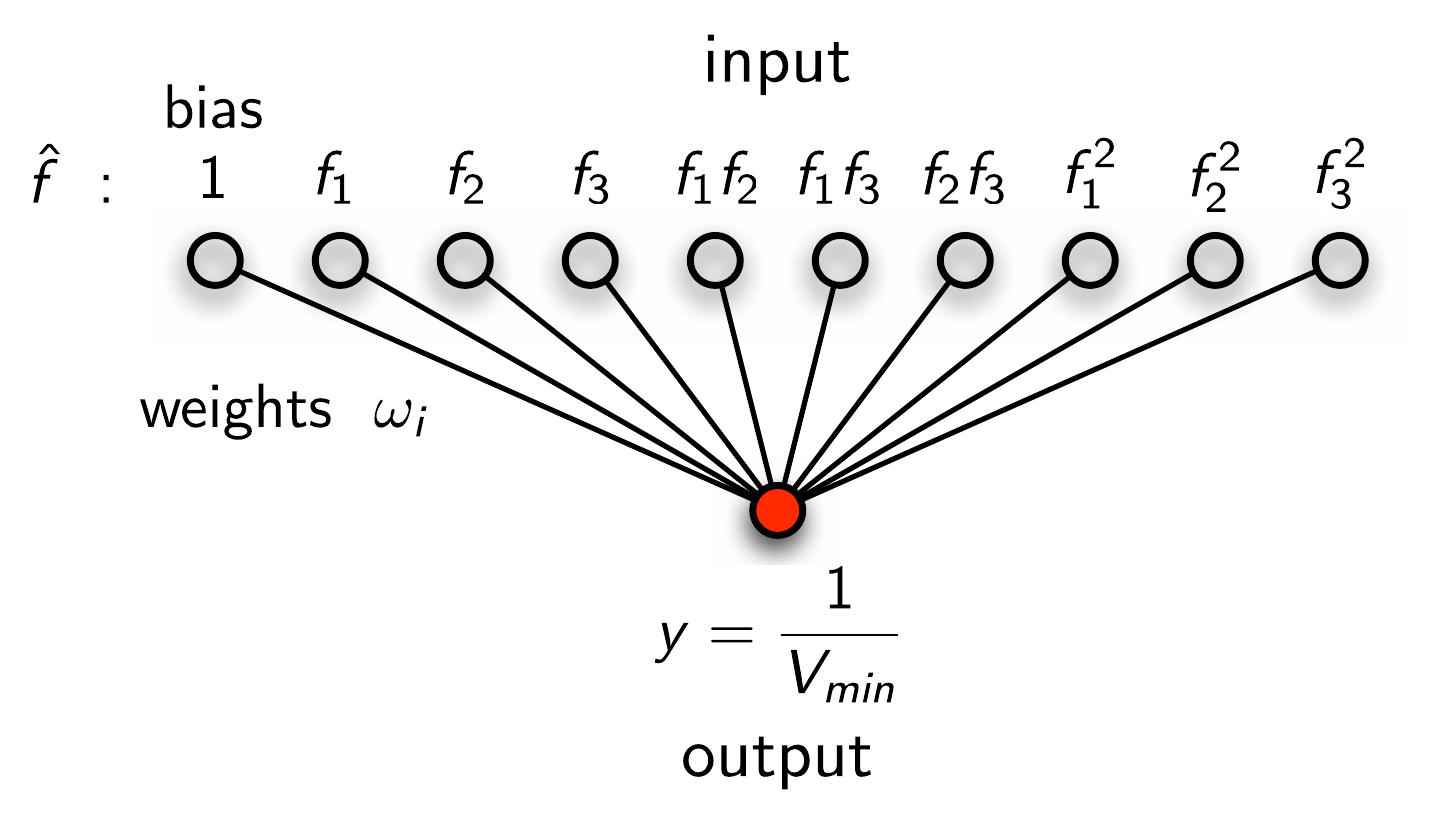}
  \caption{Illustration of the linear regression model. The features are weighted by the respective weights and added to form the output. The bias weight $w_0$ is illustrated as an additional feature, which is fixed to 1.
}
  \label{fLRnetdia}
\end{figure}

Let us first see how well we can model the data of the three features given in \req{classicalFeatures} via a simple multiple linear regression, \ie, 
\beal{es50}
y^{(n)}\sim F(f^{(n)})=\sum_{i=1}^{k_f} \omega_i\, f_i^{(n)} +\omega_0\,,
\eea
where $\omega_i$ denotes the $i$th weight, and the dataset is given by $(y,f)$, with $f$ being the features for target $y$. The weight $\omega_0$ is usually referred to as the bias and can be viewed as the weight of an additional feature fixed to 1. Furthermore, we improve the modeling non-linearly by taking order 2 combinations of the $k_f=3$ original features, which yields 
\beal{linearAnsatz}
\widehat k_f=2 k_f+\frac{k_f(k_f - 1)}{2}
\eea
new features of the form
\beal{es51}
\hat f = (f_1,\dots, f_{k_f}, f_1f_2,f_1 f_3,\dots ,f_1^2,\dots ,f_{k_f}^2)\,.
\eea
\fref{fLRnetdia} illustrates the setup of the linear regression model.

The optimization task is the usual mean least squares minimization 
\beal{es55}
\underset{\omega}{\rm argmin}\, \mathcal L\,,
\eea
where
\beal{es56}
\mathcal L=\frac{1}{N}\sum_{n=1}^N \left(y^{(n)}-F(\hat f^{(n)})\right)^2\,.
\eea
Though one can solve for $\omega_i$ exactly due to the convexity of the optimization problem, we prefer here to solve iteratively via stochastic gradient descent using the Python package Keras \cite{Keras} (with Theano \cite{Theano} backend) and the Nadam optimizer in default settings running for 5000 epochs (with batch size 1000). Note that we will use the same software stacks and settings in the following section.
The solution for the training set obtained via gradient descent reads (up to four digits)
\beal{es58}
\begin{array}{crcrcr}
\omega_1= & 1.9574, &\omega_2= &0.8522,   &\omega_3= &-0.7658,\\
\omega_4= &-0.0138, &\omega_5= &-0.0020,   &\omega_6= &-0.0104,\\
\omega_7=&- 0.0120,  &\omega_8=  &-0.0523, & \omega_9= &-0.0478,\\
& &\omega_0= &1.3637\,.& &
\end{array}
\nn\\
\eea

\begin{figure}
  \includegraphics[scale=0.55]{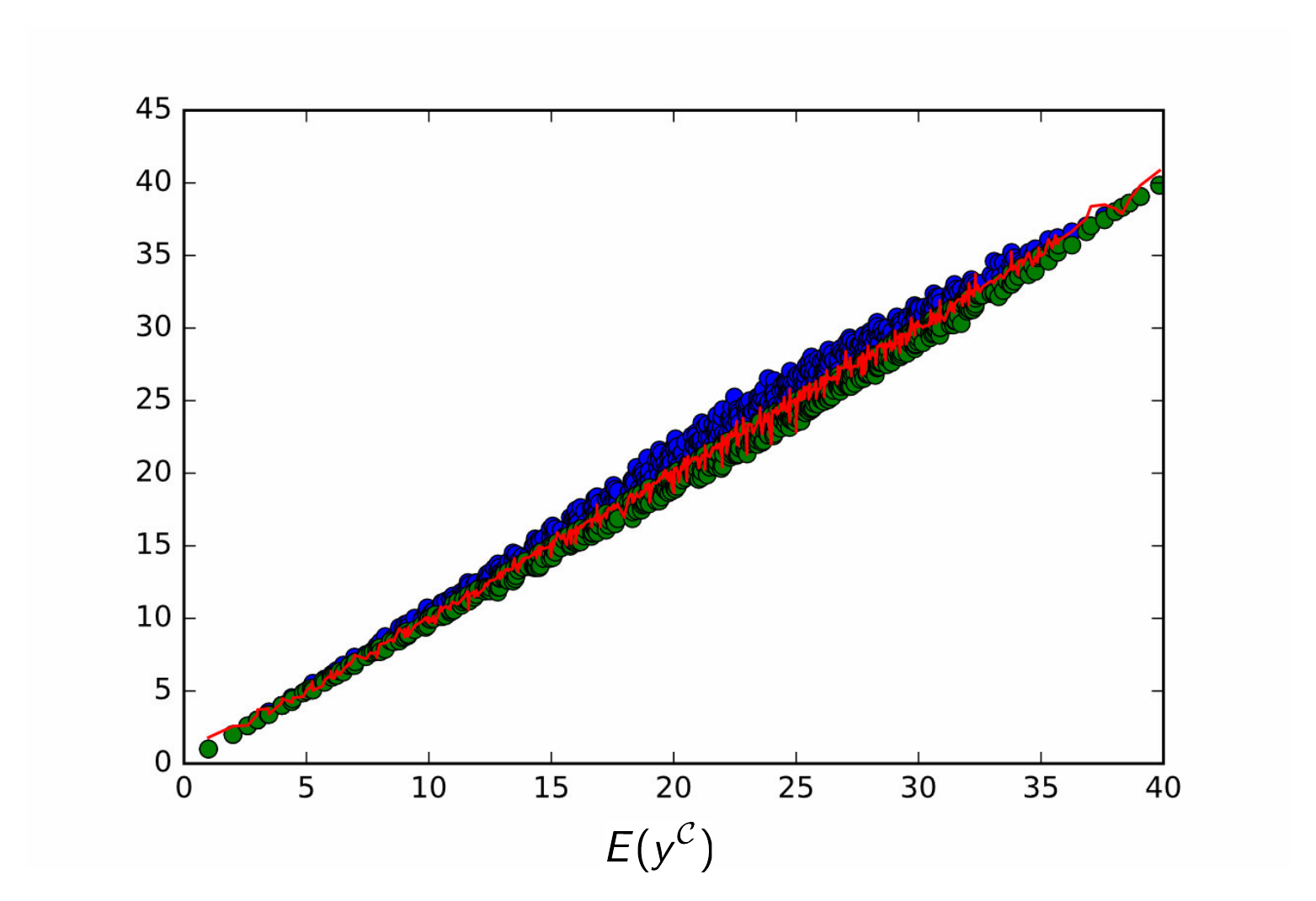}
  \caption{The $x$-axis corresponds to $E(y^{\mathcal C})$ of the 645 categories. The red curve plots the prediciton of $y$ via linear regression for the classes. The blue dots indicate the maximum $y$ taken for a class of values $\mathcal C$ and the green dots the minimum value.}
  \label{LRplot1}
\end{figure}

One should note that the features $f^{(n)}$ are not unique in our dataset. In fact, we have only 645 unique feature combinations for all the toric diagrams in our test dataset. Hence, we expect that the solution Ansatz \req{linearAnsatz} built on the three features models the statistical expectation $E(y)$ (mean). After categorizing the test dataset into 645 categories $\mathcal C_i$, we take the expectation $E(y^{\mathcal C_i})$ of the datasets in each category. 
Furthermore, we calculate the largest and smallest $y^{\mathcal C_i}$ in each category and order the categories according to their $E(y^{\mathcal C_i})$.
 Then we plot the minimum and maximum of each category as well as $F(f^{\mathcal C_i})$ against $E(y^{\mathcal C_i})$, leading to the plot in \fref{LRplot1}. 
Note that the diagonal of the figure corresponds to $E(y^{\mathcal C})$. We observe that the prediction of $F(f^{\mathcal C})$ (red curve) indeed seems to roughly approximate the mean of the $y$ values in each of the categories.

How well does this simple Ansatz actually model the true $y$, or rather the minimum volume $1/y$ of interest? We run the test set through the predictor $F$ and calculate the percentage errors (in $\times100\%$)
\beal{es60}
\epsilon^{(n)}= 1-\frac{y^{(n)}}{F(f^{(n)})}\,.
\eea
We observe a maximum error $|\epsilon_{max}|\sim 0.132$ and 
\beal{es61}
E(|\epsilon|)\sim 0.022\,,\,\,\,\,\, \sigma(|\epsilon|)\sim  0.017\,,
\eea
where $\sigma(|\epsilon|)$ denotes the standard deviation of the distribution of the absolute value of the percentage errors. We also averaged the values over three independent runs.

Hence, the expected prediction error of the minimal volume is around $2.2\%$. This means that the linear regression Ansatz already yields a surprisingly good approximator to the minimal volume. Here we should make a remark regarding the order of feature combinations that was taken in \eref{es51}. Taking order 2 combinations gives a significant improvement in reducing the error in comparison to taking the plain three features, as the extended linear regression model seems to be able to learn better the more extreme volumes at the tails of the distribution of the minimum volumes. Including in addition order 3 combinations does not seem to yield any further improvement but rather seems to lead to a worse result.

In general, in order to improve this method further, we need to introduce additional features which are able to distinguish between the individual members of a class $\mathcal C_i$. However, instead of hand-crafting new features, we try in the following section the more modern approach to let the approximator $F$ learn the appropriate features {\it itself} from the raw data.

\section{WIDE AND CNN}

\begin{figure}
  \includegraphics[scale=0.5]{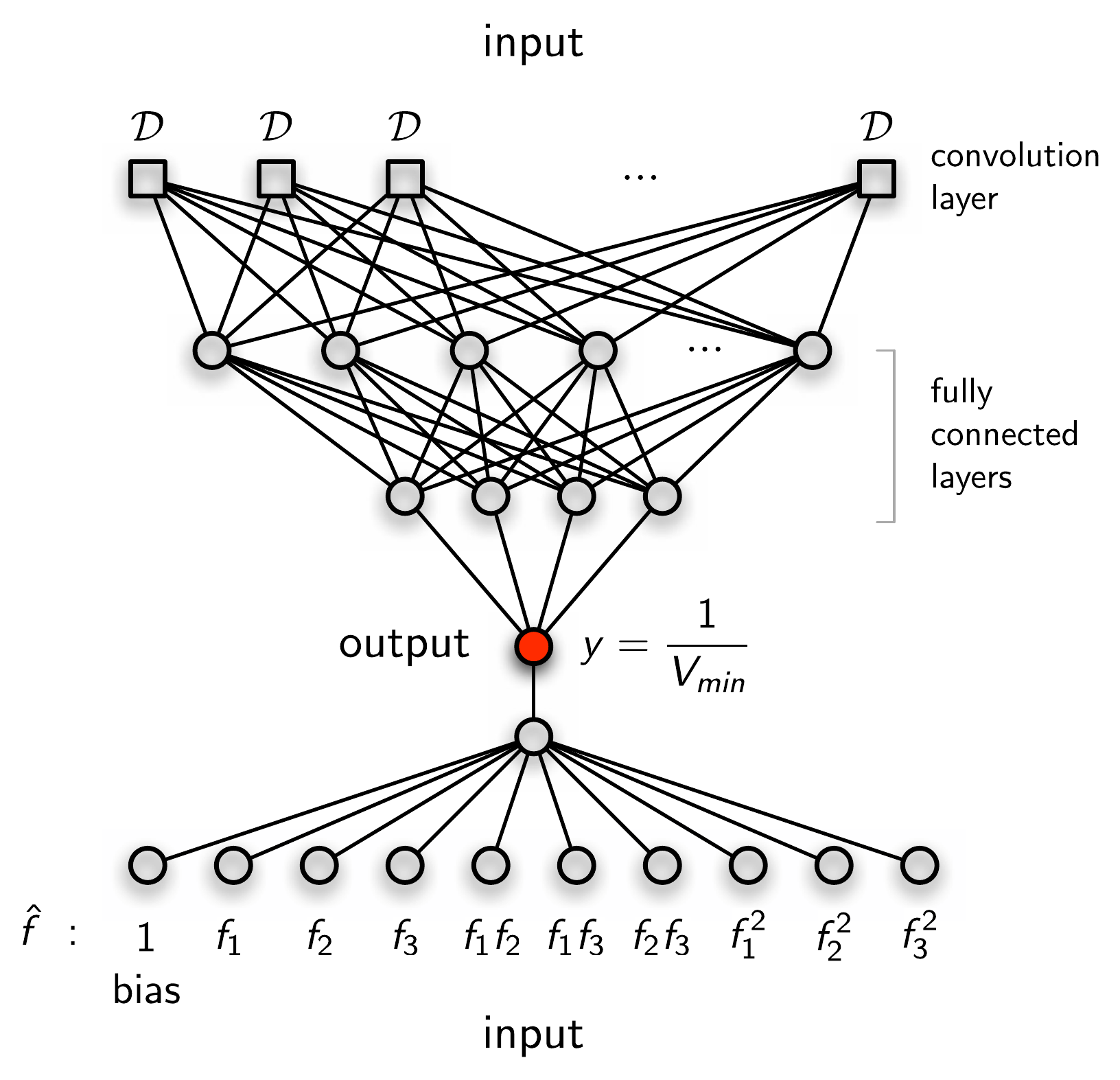}
  \caption{The wide and deep model. The toric diagram data $\mathcal D$ is fed into a convolutional layer and further processed in two fully connected layers. The outputs are linearly combined with the output of a linear regression on the features $\hat{f}$. 
}
  \label{fLRCNNnetdia}
\end{figure}

The raw data in its purest form is the toric diagram itself, hence it behooves us to ask if we can learn additional information directly from the toric diagram, in order to minimize further the prediction error. Since we treat the toric diagram as a $9\times 9$ matrix viewed as an image, the canonical computer science approach is to invoke a CNN -- the main tool for image recognition. 

We couple the linear regression, as described in the previous section, with a CNN as follows. The output of the linear regression is added via a single ReLu unit (rectified linear, $\max(x,0)$) to the outputs of the CNN, \ie, 
\beal{es62}
y^{(n)}\sim \max \left(\omega^{f}_1\, F(f^{(n)}) + \sum_{i=1}^o\omega^{f}_{2i} \mathcal\,  m_i(\mathcal D^{(n)}) +\omega^{f}_0 ,0\right)\,,
\nn\\
\eea
where $m_i$ denotes the $i$th output of the CNN and $\omega^f$ are the weights of the final layer. We use the ReLu unit for the final output because the minimum volume has to be positive. Note that this kind of setup is also known as a wide and deep model.

\begin{figure}
  \includegraphics[scale=0.5,trim={1.5cm 2cm 0cm 1cm}]{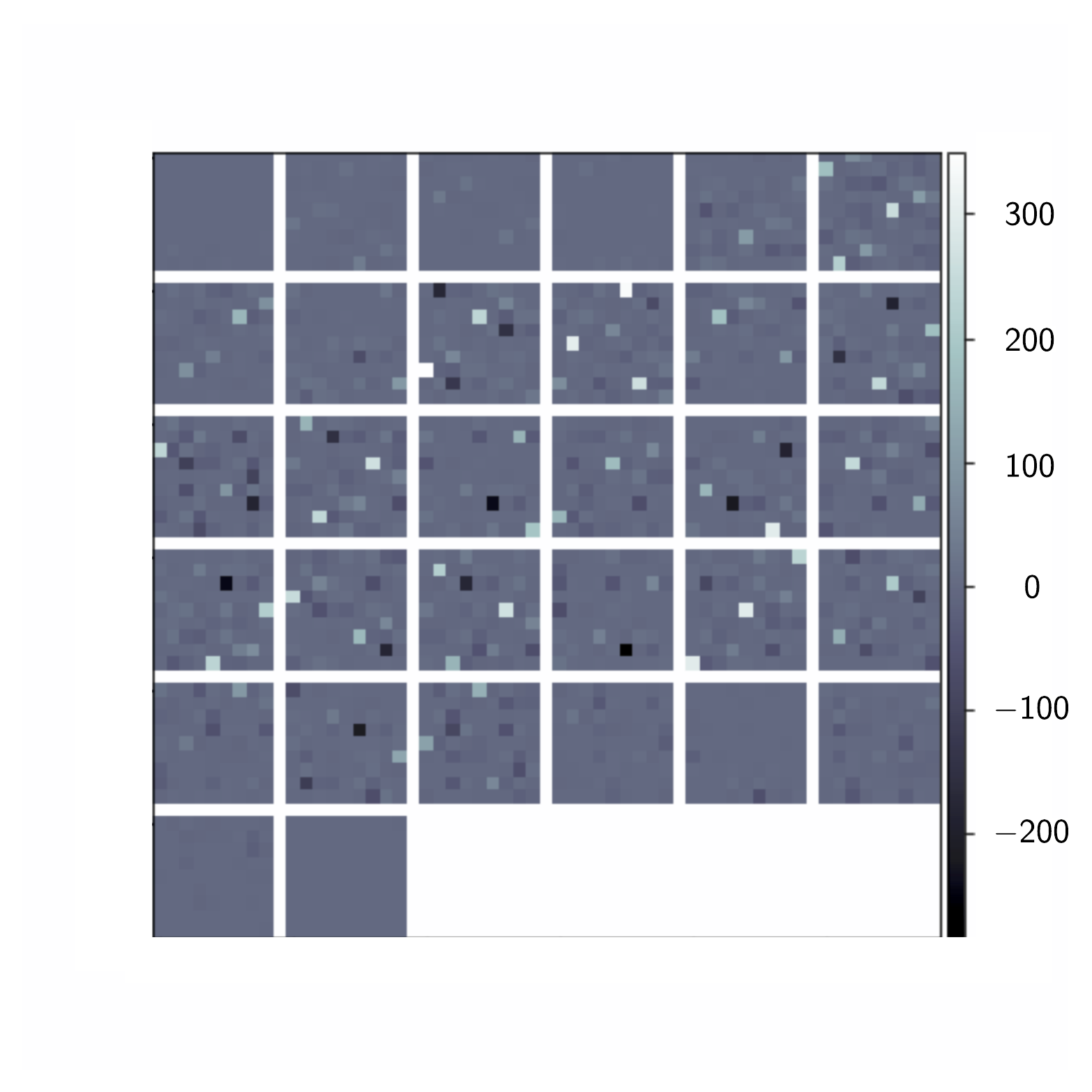}
  \caption{
  An illustration of the 32 convolutional filters acting on the average input in the CNN.
  The color shading refers to the output of the convolution layer. 
}
  \label{fCNNvisu}
\end{figure}

The CNN is setup as follows. For the input layer we take a $2d$ convolutional layer consisting of 32 filters (size $3\times 3$) and linear activation. The filters are convolved against the input and produce a $2d$ activation map of the filter. Hence, the layer learns spatially localized features of the inputs. For an illustration of the convolution layers see \fref{fCNNvisu}.
This is followed by two relatively small fully connected layers (we use sizes 12 and 4) with $\tanh$ activation. Hence we have 4 outputs $m_i$ in the CNN part. The precise network architecture is not of utmost relevance, as the results appear to be relatively stable against modifications in the number of layers, units in each layer, etc. However, generally smaller networks seem to be preferred with $\tanh$ activation functions in the dense layers. 
\fref{fLRCNNnetdia} illustrates the combined setup. 

The complete setup is trained on the train set as in the previous section via stochastic gradient descent minimizing the mean squared error. 

\begin{figure}
  \includegraphics[scale=0.55]{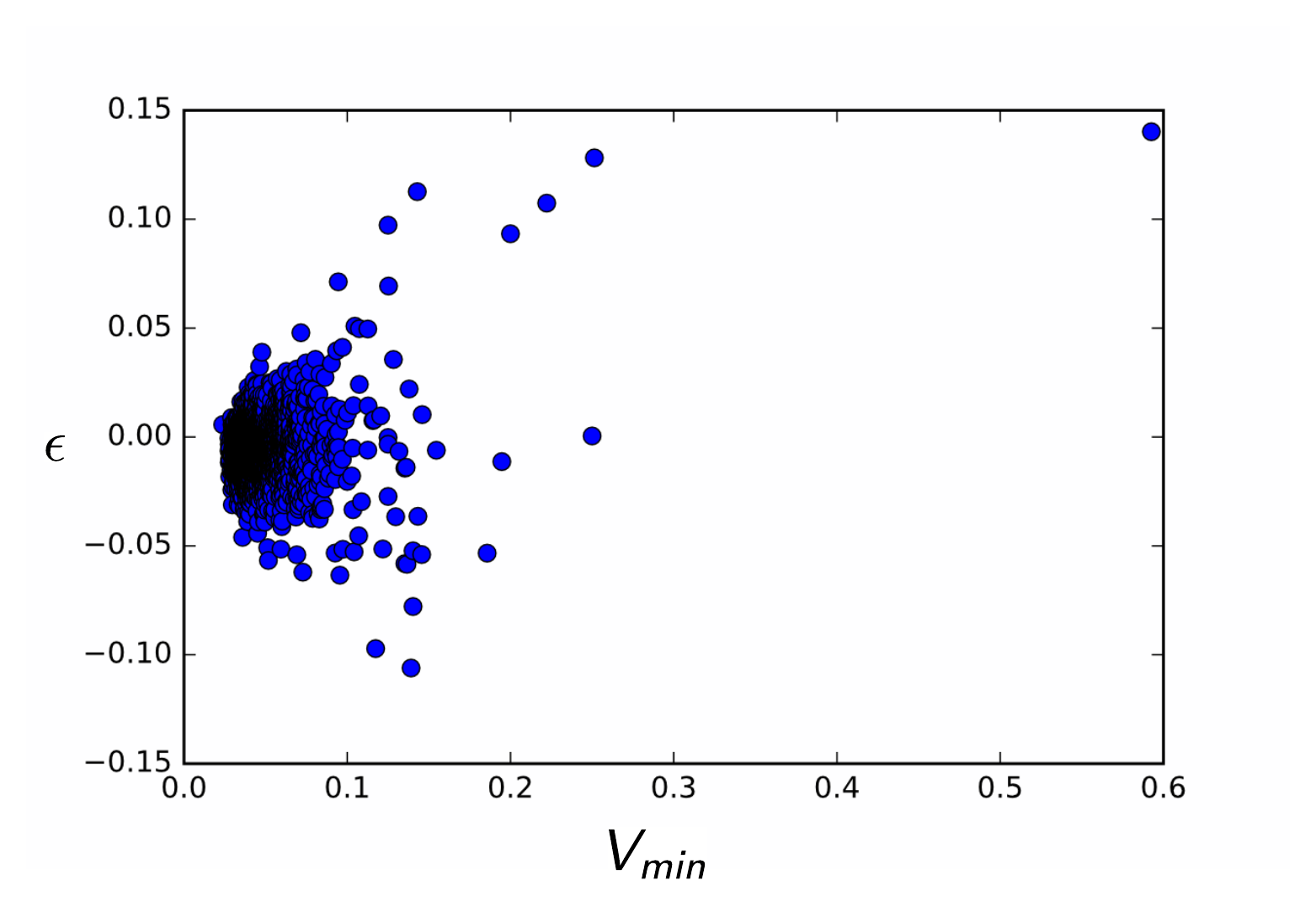}
  \caption{The $x$-axis corresponds to the minimum volume $V_{min}$ and the y-axis to the percentage error $\epsilon$ ($\times100\%$). The blue dots correspond to the errors between the prediction and should be results for the coupled linear regression and CNN.}
  \label{LRCNNerrors}
\end{figure}

Using the trained network, the prediction of the volume minima for the independent test set exhibits the following errors averaged over three independent test runs, 
$$
E(|\epsilon|) \sim  0.009 \,,\,\,\,\,\sigma(|\epsilon|) \sim 0.009 \,.
$$
Hence, the expected prediction error is below $1\%$. The maximum observed error reads $|\epsilon_{max}|\sim 0.20$. The distribution of errors for one test run is plotted in \fref{LRCNNerrors}. 
We conclude that adding the CNN yields a significant improvement in predictive power. Note that the few larger errors visible in the plot are due to the tails of the minimum volume distribution, which the model is not able to predict extremely well due to the lack of training data available at the extreme values. 

Finally, let us consider the case of using just the CNN alone, without being coupled to a linear regression branch. The used CNN is identical to the one above. 

\begin{figure}
  \includegraphics[scale=0.55]{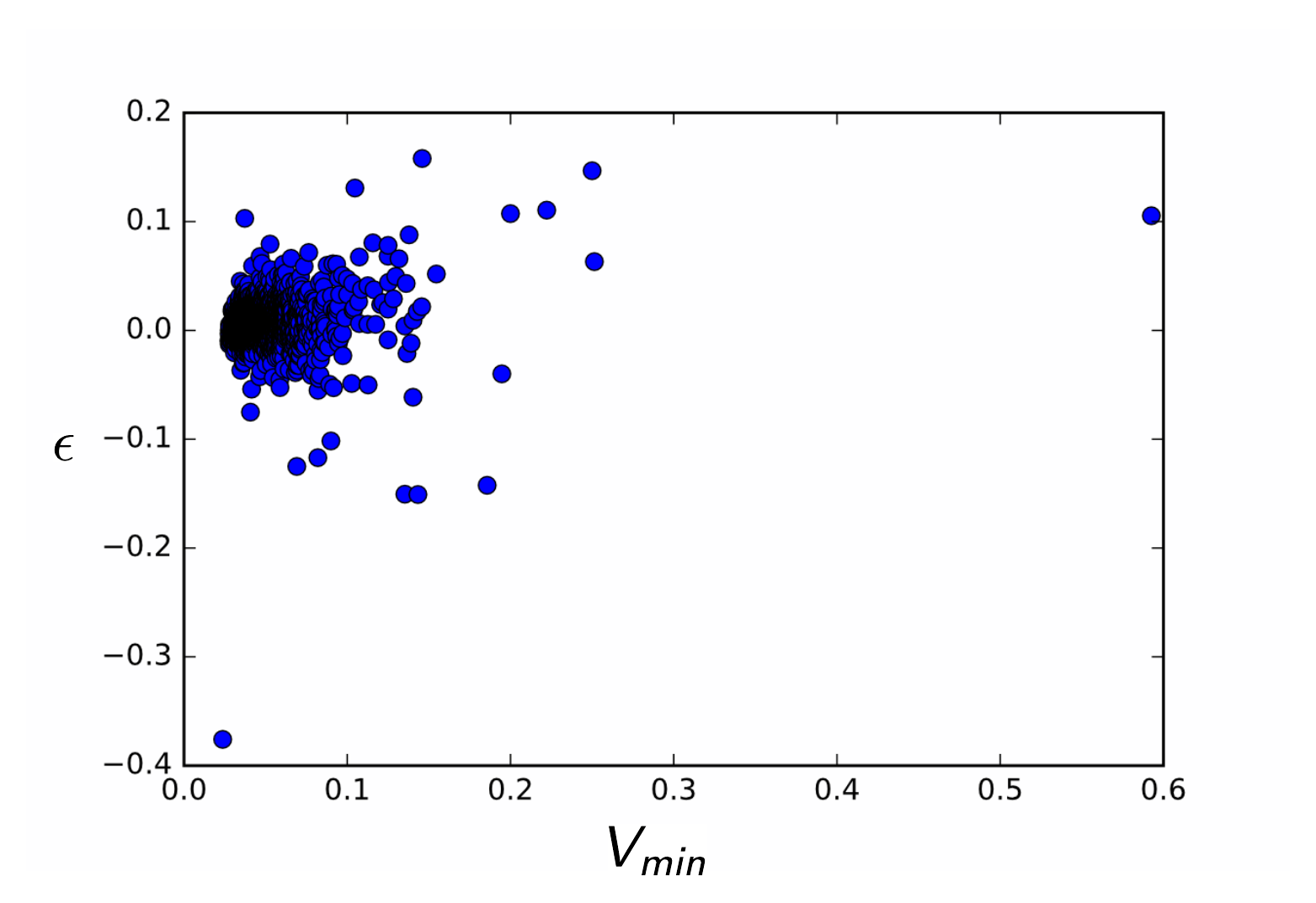}
  \caption{The $x$-axis corresponds to the minimum volume $V_{min}$ and the y-axis to the percentage error $\epsilon$ ($\times100\%$). The blue dots correspond to the errors between the prediction and should be results for the pure CNN.}
  \label{CNNerrors}
\end{figure}

We obtain on the test set
$$
E(|\epsilon|) \sim  0.010 \,,\,\,\,\,\sigma(|\epsilon|) \sim 0.014 \,,\
$$
and $|\epsilon_{max}|\sim 0.51$. Note that these values are averaged as well over three independent test runs. For illustration, the individual errors for one test run are plotted in \fref{CNNerrors}.

In general, the machine learning models have greater difficulty in learning the tails of the minimum volume distribution. 
This might be due to lack of data in this regime (the data distribution restricted to our frame is not uniform).
We observe that the combined setup of both linear regression and CNN performs better. 
This is because linear regression seems to stabilize the prediction of the tail sections of the minimum volume distribution due to its knowledge about the properties of the feature vector classes, which we discussed in the previous section.

\section{Summary and Outlook}

In this work, we have demonstrated that machine learning techniques, in particular neural networks, can be a useful addition to the toolbox of researchers tackling formal questions in mathematics and physics. 
A necessary condition for using machine learning techniques is that at least some aspect of the question can be translated to a data science problem. 

This work studied whether the minimum volume of Sasaki-Einstein base manifolds for toric Calabi-Yau 3-folds can be directly computed from topological quantities originating from toric geometry, replacing the usual minimization procedure that is necessary to identify the minimum volume. 
This question has aspects of a data science problem, \ie, out of known topological data and the corresponding minimal volumes, can we (or rather the machine) learn a mapping between these quantities (that is, find an approximate functional relation)?

The answer seems to be affirmative. Even taking for the machine learning model just a linear combination of order 2 combinations of numbers for different characteristic vertices in the toric diagrams of the Calabi-Yau 3-folds, yields already a good universal approximation to the minimal volume of the corresponding Sasaki-Einstein manifolds. 
In addition, a CNN model, which learns new kinds of features of the toric diagram, further improves predictive power for the volume minimum. 

It is surprising to see that the simple setups we are considering are able to predict the minimum volumes relatively well, effectively showing that the minimum volume is encoded in the toric diagrams of the Calabi-Yau 3-folds. 
This is an indication that the procedure of volume minimization can be avoided. In fact, we show that the volume minimum can explicitly be computed from the toric data, by using an underlying functional relation, which we have approximated in this work.

With this analysis, we have shown a working example of how a machine learning model can identify functional relationships between mathematically and physically interesting quantities in cases where such functional relationships were not known before. 

Note that it would be interesting to refine our analysis by increasing the dataset of toric Calabi-Yau's in such a way that the minimum volume distribution is more uniform. 
Furthermore, averaging over more test runs as well as increasing the rounding precision for irrational values for minimum volumes would be important improvements that we leave for future work. 

We believe that there are other suitable problems which can be approached from a data science perspective, similar to what we have done in this work for the minimum volume of toric Calabi-Yau 3-folds. 
Approaching problems in this way might yield some novel insights, as well as hints to hidden and unexpected relations between physically and mathematically relevant quantities that have not been observed before.

For example, large datasets of both physical and mathematical significance exist in the context of $4d$ $\mathcal{N}=1$ theories related to toric Calabi-Yau 3-folds \cite{Klebanov:1998hh,Hanany:1997tb,Hanany:1998it,Franco:2005rj}, as well as to a recently discovered new class of $2d$ $(0,2)$ theories related to toric Calabi-Yau 4-folds \cite{Franco:2015tya}. Furthermore, interesting rich datasets exist in relation to so called complete intersection Calabi-Yaus (CICYs) \cite{Candelas:1987kf} characterized by configuration matrices that can be taken as inputs for machine learning models. Finally, large datasets exist in relation to hyperbolic 3-manifolds related to knots \cite{jones1985polynomial}, which may exhibit hidden structures that could be discovered using again machine learning techniques. In a future work \cite{KreflSeongYau2017}, we hope to shed light on these interesting problems.

\acknowledgments
We thank S.-T. Yau for related discussions, collaborations and encouragement to pursue this project. R.-K. S. also thanks the CERN Theory Group, where this project was initiated and the Center for Mathematical Sciences and Applications at Harvard University and the Yau Mathematical Sciences Center at Tsinghua University, for their hospitality.


\bibliographystyle{jhep}
\bibliography{mybib}

%
%

\end{document}

%% file: pref.tex
\newcommand{\be}{\begin{equation}}
\newcommand{\ee}{\end{equation}}
\newcommand{\beq}{\begin{equation}}
\newcommand{\beql}[1]{\begin{equation}\label{#1}}
\newcommand{\eeq}{\end{equation}}
\newcommand{\ba}{\begin{array}}
\newcommand{\ea}{\end{array}}
\newcommand{\bea}{\begin{eqnarray}}
\newcommand{\beal}[1]{\begin{eqnarray}\label{#1}}
\newcommand{\eea}{\end{eqnarray}}
\newcommand{\ben}{\begin{enumerate}}
\newcommand{\een}{\end{enumerate}}
\newcommand{\bean}{\begin{eqnarray*}}
\newcommand{\eean}{\end{eqnarray*}}
\newcommand{\eref}[1]{(\ref{#1})}

\newcommand{\nn}{\nonumber}

\newcommand{\fref}[1]{Figure \ref{#1}}
\newcommand{\btab}[1]{\begin{tabular}{#1}}
\newcommand{\etab}{\end{tabular}}

\newcommand{\comment}[1]{}

\newcommand{\ud}{\mathrm{d}}

\newcommand{\qed}{\nobreak \ifvmode \relax \else
      \ifdim\lastskip<1.5em \hskip-\lastskip
      \hskip1.5em plus0em minus0.5em \fi \nobreak
      \vrule height0.75em width0.5em depth0.25em\fi}

\definecolor{darkspringgreen}{rgb}{0.09, 0.45, 0.27}
\definecolor{forestgreen}{rgb}{0.13, 0.55, 0.13}